\documentclass[aps,twocolumn,prd,showpacs,nofootinbib,showkeys,superscriptaddress,preprintnumbers]{revtex4-1}

\usepackage{lineno,hyperref}
\modulolinenumbers[5]
\usepackage{graphicx,floatflt}
\usepackage{amsmath}
\usepackage{amstext}
\usepackage{amssymb}
\usepackage{epsfig}
\usepackage[usenames]{color}
\usepackage[dvipsnames]{xcolor}

\newcommand{\dd}{\mathrm{d}} 

\newcommand{\rr}{\mathrm}

\newcommand{\sun}{\odot}

\newcommand{\mPBH}{m_{\rr{\rm PBH}}}
\newcommand{\deltaloc}{\delta_{\rr{\rm PBH}}^{\rr{loc.}} }

\newcommand{\be}{\begin{equation}}
\newcommand{\ee}{\end{equation}}
\newcommand{\ba}{\begin{eqnarray}}
\newcommand{\ea}{\end{eqnarray}}
 
\begin{document}

\preprint{TTK-16-10, IFT-UAM/CSIC-16-027}

\title{The clustering of massive Primordial Black Holes as Dark Matter: \\ measuring their mass distribution with Advanced LIGO}

\author{S\'ebastien Clesse}
\email{clesse@physik.rwth-aachen.de}
\affiliation{Institute for Theoretical Particle Physics and Cosmology (TTK), RWTH Aachen University, D-52056 Aachen, Germany}

\author{Juan Garc\'ia-Bellido} \email{juan.garciabellido@uam.es} 
\affiliation{Instituto de F\'isica Te\'orica UAM-CSIC, Universidad
Auton\'oma de Madrid, Cantoblanco, 28049 Madrid, Spain}



\date{\today}

\begin{abstract}
The recent detection by Advanced LIGO of gravitational waves (GW) from the merging of a binary black hole system sets new limits on the merging rates of massive primordial black holes (PBH) that could be a significant fraction or even the totality of the dark matter in the Universe.   aLIGO  opens the way to the determination of the distribution and clustering of such massive PBH.   If PBH clusters have a similar density to the one observed in ultra-faint dwarf galaxies, we find merging rates comparable to aLIGO expectations.  Massive PBH dark matter predicts the existence of thousands of those dwarf galaxies where star formation is unlikely because of gas accretion onto PBH, which would possibly provide a solution to the missing satellite and too-big-to-fail problems.   Finally, we study the possibility of using aLIGO and future GW antennas to measure the abundance and mass distribution of PBH in the range [5 - 200] $M_\odot$ to 10\% accuracy. 
\end{abstract}
\pacs{98.80.Cq}
\maketitle

Understanding the nature of Dark Matter (DM), accounting for about one third of the energy density of the Universe, is one of the most important challenges in cosmology nowadays (for a review, see e.g.~\cite{Bertone:2010zza}).
A popular hypothesis is that DM is composed of Weakly Interacting Massive Particles (WIMP's).  However, in the absence of a clear signal from direct or indirect WIMP interactions, possible alternatives should be considered.  

For instance, DM could be composed partially or totally in the form of Primordial Black Holes (PBH)~\cite{Carr:1974nx,Carr:1976zz,Carr:1975qj,Meszaros:1975ef,GarciaBellido:1996qt,Khlopov:2008qy,Frampton:2010sw,Belotsky:2014kca}.  These could have formed in the early universe due to the collapse of large density fluctuations, e.g. induced by a waterfall phase during inflation~\cite{GarciaBellido:1996qt,Clesse:2015wea,Lyth:2011kj,Bugaev:2011wy} , by a first-order phase transition~\cite{Jedamzik:1999am} or in some curvaton scenarios~\cite{Kohri:2012yw,Kawasaki:2012wr,Bugaev:2013vba}.   Like a WIMP, a PBH is non-relativistic and effectively collisionless and is thus a perfect DM candidate.  PBHs must be heavy enough not to evaporate in a time shorter than the age of the Universe, which is fulfilled if their mass is $\mPBH \gtrsim 5 \times 10^{11}$ kg~\cite{Carr:1976zz,Carr:2009jm}.  A noticeable exception is the possibility that PBH form stable relics with Planck-like mass~\cite{Carr:2016drx} after their evaporation.  Very stringent constraints have been set on their abundances, from various observations:  if  $\mPBH \lesssim 7 \times 10^{12}$~kg, the gamma-ray radiation due to PBH evaporation should have been detected by EGRET and FERMI~\cite{Carr:2009jm};  within the range $5 \times 10^{14} - 10^{17}$~kg, they should have been detected by FERMI through the gravitational femto-lensing of gamma-ray bursts~\cite{Barnacka:2012bm};  for $10^{15} < \mPBH < 10^{21}$~kg PBHs should have destroyed neutron stars in globular clusters~\cite{Capela:2013yf};  the absence of microlensing events of stars in the Magellanic clouds exclude large abundances of PBHs within the range $10^{23} - 10^{31}$~kg \cite{Tisserand:2006zx,Alcock:1998fx,Griest:2013aaa}, although such constraints are model dependent~\cite{Hawkins:2015uja,Sasaki:2016jop}.   Their abundance in the early Universe is also well constrained by the absence of important spectral distortions of the CMB black-body spectrum, which excludes PBH as dark matter if $\mPBH \gtrsim  1 M_\odot $~\cite{Ricotti:2007au}.   This last constraint closes the range of possible PBH masses, therefore most people often considers the model as ruled out.  

However, as was pointed out recently in Ref.~\cite{Clesse:2015wea}, the merging of PBHs could have been very efficient in the early Universe, such that initially substellar mass black holes, passing the CMB distortion constraints, could have grown by several orders of magnitude, enough to evade the most stringent microlensing constraints.   In this way, the galactic halo would be populated by a large number of massive PBHs, which is consistent with the recent observation of numerous BH candidates in the central region of Andromeda and nearby galaxies~\cite{Barnard:2014afa,Barnard:2013nqa,Barnard:2013dea,Barnard:2012tn,Barnard:2011pv}. Moreover, recent analysis of the gamma-ray excess seen by Fermi-LAT towards the Galactic Center finds evidence for a population of unresolved point sources~\cite{Bartels:2015aea,Lee:2015fea} which could be, together with the 30\% unidentified point sources of the 3FGL catalog~\cite{Acero:2015gva}, the tip of the iceberg of the PBH distribution of Dark Matter.

In addition, in the case of a broad PBH mass spectrum, covering a few orders of magnitude, the high-mass tail of the distribution provides a subdominant number of very massive PBHs, which could quick-start structure formation and, in particular, are good candidates for the seeds of the Super-Massive Black Holes (SMBH) observed at the center of galaxies\footnote{Such a scenario is well-constrained by the absence of CMB distortions~\cite{Kohri:2014lza} but is allowed if the power spectrum of curvature perturbations is only enhanced on smaller scales than the ones relevant for $\mu$-type distortions, as in the model proposed in Ref.~\cite{Clesse:2015wea}.}, as well as for the Intermediate Mass Black Holes (IMBH) in Globular Clusters~\cite{Kruijssen:2013cna}. The recent observation of one of those IMBH in the central region of the Milky-Way~\cite{2016ApJ...816L...7O} could be a hint in favor of abundant IMBHs, beyond what is expected via stellar evolution. Moreover, massive PBH could be responsible for the observed ultra-luminous X-ray sources~\cite{2015MNRAS.448.1893M,Dewangan:2005mb,Madhusudhan:2005zj,Liu:2013jwd,Bachetti:2014qsa}.  

In this \textit{letter}, we examine the possibility that the $\sim 30 M_\odot$ BH merger at the origin of the first direct detection of gravitational waves by Advanced LIGO~\cite{Abbott:2016blz} has a Dark Matter origin in the form of PBHs.  More generally, we explore how the new bounds set by aLIGO on the rate of massive BH merging~\cite{Abbott:2016nhf} can be satisfied by a massive PBH-DM model.   Two cases are distinguished:  first, the PBHs are uniformly distributed in galactic halos and follow Einasto or Navarro-Frenk-White (NFW) profiles; second, PBHs are clustered in compact sub-halos.   In both cases we compute the merger rate and 
evaluate the typical size and density of those PBH sub-halos leading to a 
rate within the range of $2-400 \  \rr{yr}^{-1} \rr{Gpc}^{-3}$ inferred from aLIGO observations~\cite{Abbott:2016nhf}.  

Upon completion of our work, S. Bird et al. released a similar analysis~\cite{Bird:2016dcv}~\footnote{Soon after the first version of the present letter, another similar analysis was released by Sasaki et al. \cite{Sasaki:2016jop} with similar conclusions.}.  Here we extend their claim that aLIGO could have detected PBH-DM to the case that black holes are clustered: (i) rather than a scale-invariant spectrum 
we use the predicted peak in the matter power spectrum on small scales and assume that PBH could have already clustered in the early Universe; (ii) we consider the case of a broad mass distribution of the PBH spectrum, rather than a single-mass ``monochromatic" spectrum;  (iii) we provide a mechanism for generating such initially clustered PBH in the context of inflation with a mild-waterfall phase; (iv) we discuss the implications of our results for the missing-satellites and too-big-to-fail problems~\cite{Bertone:2010zza}, which could be solved naturally if PBH-DM are mostly concentrated in faint dwarf galaxies with high mass-to-light ratios.

\textit{Uniform distribution in the Milky-Way Halo:}  
We have assumed that the galactic DM density follows an Einasto profile~\cite{1989A&A...223...89E},
\be
\rho(r) = \frac{\rho_{-2} }{\rr e^{2 n \left[  (r/ r_{-2})^{1/n} - 1 \right]}}
\ee
where $r_{-2} $ is the radius at which the logarithmic slope of the profile equals $-2$, and  the density parameter $\rho_{-2} \equiv \rho(r_{-2})$.
These are equivalent to the radius $R_{\rr s}$ and density $\rho_{\rr s}/4$ of the common NFW profile, respectively. Typical values from DM simulations for massive halos, like the Milky-Way halo for which $r_{-2} \approx 20$ kpc, give $ 4 \lesssim n \lesssim 7$.  The case $n=4$ is considered below, but it has been checked that different values only affect marginally the results.  
We also assume that PBHs follow a Maxwellian distribution with  average velocity $\bar v = 200$ km/s.   Varying those parameters in a reasonable way would have a rather limited impact on the merging rate, so for simplicity they are kept constant throughout the paper.  If PBH are uniformly distributed within the galactic halo, i.e. they are not clustered, the typical distance between two BHs 
goes from a few to tens of parsecs, for PBH masses in the range $ 1 \lesssim m_{\rr{\rm PBH}} / M_\odot  \lesssim 100$. 

\begin{figure}
\begin{center}
\includegraphics[scale=0.95]{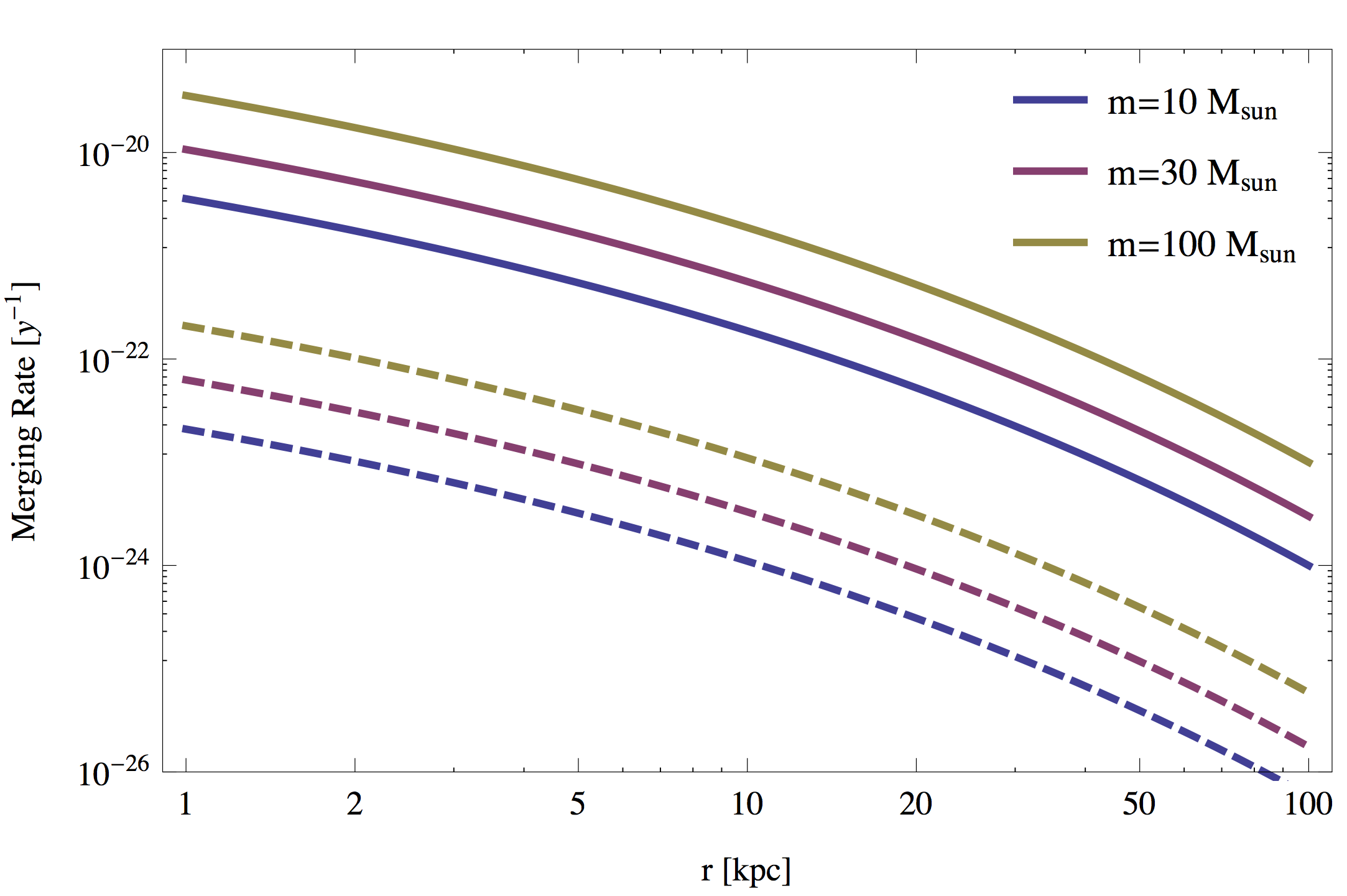}
\caption{PBH capture rate (solid lines) and direct merging rate (dashed lines) in the Milky-Way, for a massive PBH-DM model uniformly distributed with an Einasto profile, within the Milky-Way halo, and for several values of the PBH mass. }
\label{fig:galhalo}
\end{center}
\end{figure}

In order to calculate the merging rate, we have referred to Refs.~\cite{1989ApJ...343..725Q,Mouri:2002mc} where the capture cross-section $\sigma^{\rr{capt}}$ of two black-holes in dense clusters has been calculated.  Two encountering BHs become bounded to each other if the energy lost in the form of gravitational waves is of the order of the kinetic energy.   Once gravitationally bounded to each other, the two BH quickly merge in less than a million years. In the Newtonian approximation, this capture rate $\tau^{\rr{capt}}_{\rr{\rm PBH}}  \equiv n_{\rr{\rm PBH}}~\bar v ~\sigma_{\rr{\rm PBH}}^{\rr{capt}}$ of a BH of mass $m_{\rr A}$ by a BH of mass $m_{\rr B}$ is given by~\cite{Mouri:2002mc}
\ba 
\tau_{\rr{\rm PBH}}^{\rr{capt}} & = & (2 \pi) \ n_{\rr{\rm PBH}}(m_{\rr A}) \bar v \  \left( \frac{85 \pi}{6 \sqrt 2} \right)^{2/7} \nonumber \\  
& & \times \frac{G^2 (m_{\rr A} + m_{\rr B})^{10/7}  (m_{\rr A} m_{\rr B})^{2/7}  c^{18/7}} {c^4 v_{\rr{rel}}^{18/7}}    \label{eq:capture_rate}
\ea
where $n_{\rr{\rm PBH}} $ is the PBH number density and $v_{\rr{rel}}$ is the relative velocity of the two BHs, which we take equal to $\bar v$.  Since the cross-section is much larger than the BH surface area, the Newtonian approximation is valid.   The capture rate can be compared to the direct merging rate $\tau_{\rr{\rm PBH}}^{\rr{merg}}$, which was derived in the Newtonian approximation in Refs.~\cite{Kouvaris:2007ay,Capela:2013yf} in the context of WIMP-neutron star and PBH-neutron star collisions respectively, assuming that two PBH merge if the closest distance between them is smaller than the Schwarzschild radius $R_{\rr{\rm PBH}} = 2 G m_{\rr{\rm PBH}} / c^2$.  This rate is given by 
\be  \label{eq:merging_rate}
\tau_{\rr{\rm PBH}}^{\rr{merg}} = n_{\rr{\rm PBH}}(m_A) \left( \frac{3}{2 \pi \bar v^2} \right)^{3/2} \frac{8 \pi^2 G} {3}   m_{B} R_{B} \bar v^2 ~.
\ee
The direct merging and capture rates have the same mass dependence and are represented on Fig.~\ref{fig:galhalo} as a function of the radial distance to the galactic center, for $m_{\rr A} = m_{\rr B} = 30 M_\odot$.   The capture rate is $\sim$ 170 times larger than the direct merging rate.  General relativistic corrections could enhance the direct merging rate by a factor of a few, as noticed in Ref.~\cite{Kouvaris:2007ay}, but merging through direct collisions can be considered as a subdominant process.  Individual capture rates are found to be lower than $10^{-19} \rr{yr}^{-1}$.    Integrating over all the PBHs inside the galactic halo, one gets the total rate in our galaxy, $\tau_{\rr{gal}} \approx 5 \times 10^{-12} \mathrm{yr}^{-1}$.   This rate is comparable to the probability of star collisions within the galactic disk.  It is very low, and is found to be independent of the mass of PBHs.   The exact shape of the density profile is of little importance, similar results being obtained for the common NFW profile, with variations in the merging rate not exceeding a few percent.  Therefore, one can conclude that if PBH are uniformly distributed in the halos of galaxies, the model passes all the present and future constraints from gravitational wave experiments.   This also confirms that PBHs are effectively collisionless, as expected for a good dark matter candidate.
We have considered above the merging rate within our galaxy and extrapolated it to all possible galaxies in our local Universe up to 450 Mpc. It is clear that the rate of mergers from a uniformly distributed PBH population in the local universe is not enough to account for the LIGO observations. These could only come from high density regions with large mass-to-light ratios where PBH could be highly concentrated, like Dwarf Spheroidals or Globular clusters in our local cosmological neighborhood up to $z\sim0.12$.

\textit{Clustering in sub-halos:}
Depending on the process of formation, as well as on the evolution of cosmic inhomogeneities, it is possible that nowadays massive PBHs are regrouped in dense clusters, whose size could range from a few parsecs to a few hundreds.  Early clustering of PBHs is expected e.g. in the scenario proposed in Ref.~\cite{Clesse:2015wea}, in which quantum diffusion close to a tachyonic instability in hybrid inflation leads to different perturbation dynamics during the subsequent mild-waterfall phase~\cite{Clesse:2010iz,Clesse:2013jra}.   As a result, the formation of PBHs is expected to occur in localized regions during the radiation era, whereas on CMB scales, the primordial spectrum is unchanged and the level of non-gaussianity and isocurvature modes as expected for a slow-roll single-field model, thus evading the constraints of \cite{Young:2015kda,Bartels:2015aea}~\footnote{ Another PBH clustering scenario was considered in~\cite{Chisholm:2005vm,Dokuchaev:2008hz} but only applies to abundances less than the dark matter~\cite{Young:2015kda,Bartels:2015aea}}. Another possibility is that PBH have clustered during the cosmic history, following the non-linear evolution of DM fluctuations. 

A possible signature of such a clustered distribution could come from caustic crossing of microlensing events of massive PBH, where superimposed on the principal light amplification curve appear distinctive caustics due to smaller orbiting bodies. One such event in fact was measured by the MACHO collaboration~\cite{Alcock:1998re}.

We shall show here that a PBH-DM model with important clustering could reach the large merging rates observed by aLIGO. For this purpose one can use directly Eq.~(\ref{eq:capture_rate}) with a local number density $ n_{\rr{\rm PBH}} = \deltaloc  \times \bar \rho_{\rm DM} / m_{\rm PBH} $, with $\bar \rho_{\rm DM} = \Omega_{\rm DM} \rho_{\rr{cr}}$ being the mean cosmological DM density today, and $\deltaloc = \rho_{\rm PBH}/\bar\rho_{\rm PBH}-1\simeq f_{\rm DM}^{-1}\rho_{\rm PBH}/\bar\rho_{\rm DM}\gg1$ the local density contrast in PBH.  One finds that the individual capture (merging) rate 
is given by
\be
\tau_{\rr{\rm PBH}}  \simeq   10^{-28} ~  \deltaloc \left( \frac{\mPBH}{M_\odot} \right) \rr{yr}^{-1} \,,
\ee
and that the total rate per Gpc$^3$ is
\be
\tau_{\rm tot}  \simeq   7 \times 10^{-9} ~f_{\rm DM} ~ \deltaloc \ \rr{yr}^{-1} \rr{Gpc}^{-3} \,,
\ee
independently of the PBH mass\footnote{A factor $ f_{\rm DM} $ is included to consider the case PBHs account for only a fraction of dark matter.  The case $ f_{\rm DM} = 1 $ is discussed thereafter, but since it simply rescales the total merger rate it is straightforward to extend our results to the case $ f_{\rm DM} < 1$.}. Therefore, a large local density contrast of PBH with respect to the cosmological DM density, $\deltaloc \sim 10^{9}  - 10^{10}$, is typically required to produce a few to tens of events per year, inside the range of Advanced LIGO~\cite{TheLIGOScientific:2016wyq}. For $\mPBH \simeq 30 M_\odot$, this corresponds to a local number density of $\sim 1$ ($\sim 10$) PBHs per cubic parsec.  The PBH density is comparable to the one of DM in globular clusters, as well as to that of DM-dominated - with a mass-to-light ratios approaching $\sim 1000$ -  ultra-faint dwarf galaxies detected by Keck/DEIMOS~\cite{Martin:2007ic,Simon:2007dq}, and is not far from the density inside the compact clusters observed around a nearby galaxy~\cite{2015ApJ...805...65T}.   The existence of hundreds of those ultra-faint dwarf satellite galaxies has been recently inferred by the DES collaboration~\cite{Drlica-Wagner:2015ufc}.  Interestingly, the Keck ultra-faint galaxies~\cite{Martin:2007ic,Simon:2007dq} have a total mass of $10^6 - 10^7 M_\odot$ for radii ranging from a few tens to a few hundreds of parsecs, which fits well with the required density to reproduce the merger rate inferred by Advanced LIGO.  With such densities, we predict the existence of hundreds to thousands of those ultra-faint dwarf satellites around each galaxy.   

\textit{Missing satellite problem:} N-body simulations of the $\Lambda$CDM problem predict that there should exist numerous of such dwarf satellite galaxies, that are not observed in visible light, which is referred to as the missing satellite problem~\cite{Bertone:2010zza}.  Another problem of the $\Lambda$CDM model, referred as the too-big-to-fail problem, is that large sub-structures, if present, lead to massive star formation and therefore should have been detected already~\cite{Bertone:2010zza}. 

\begin{figure}[ht]
    \begin{center} 
        \includegraphics[width=7cm]{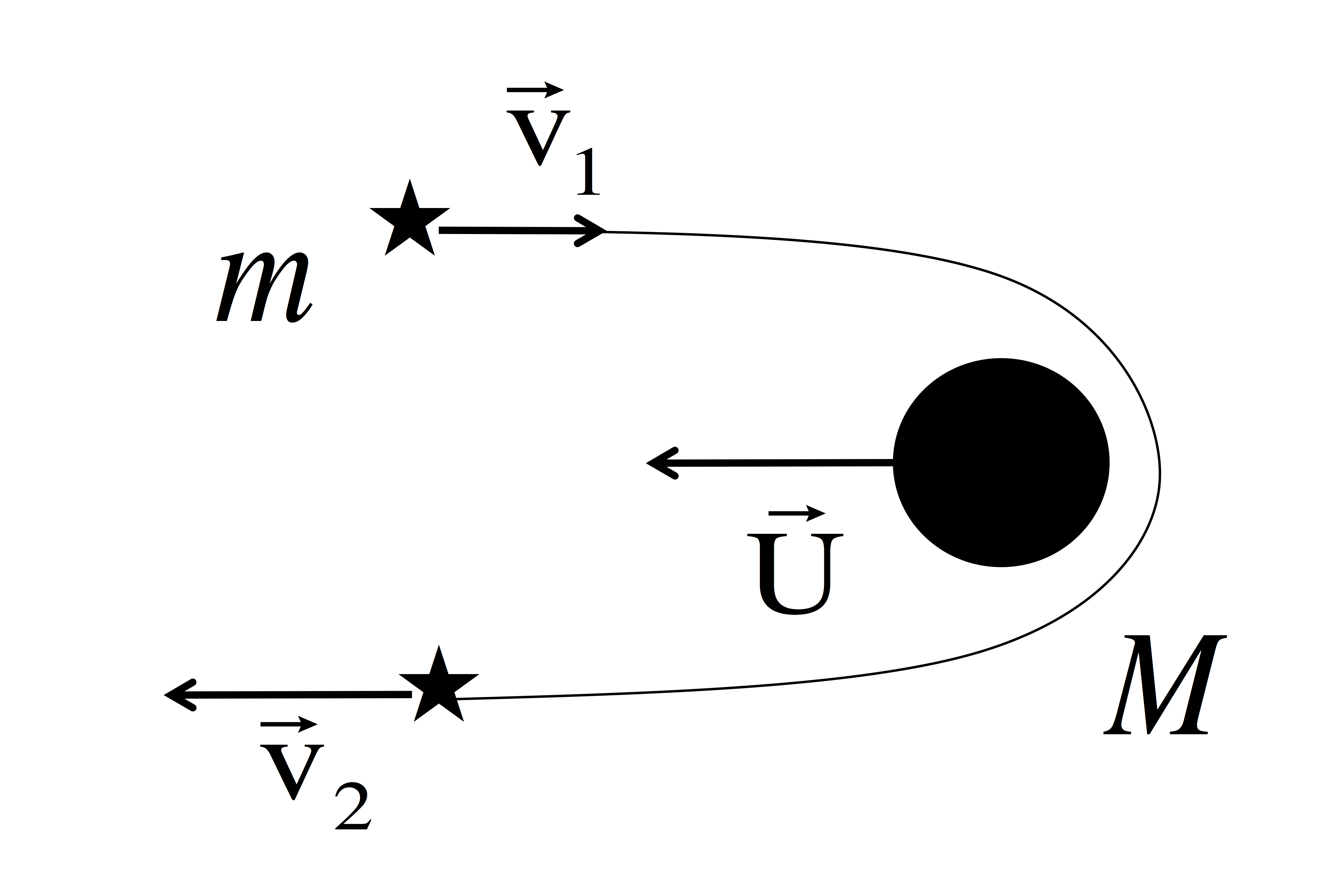}
    \end{center}
    \caption{The gravitational slingshot effect converts a small peculiar velocity $\vec{v}_1$ of a star of mass $~m~$ into a larger than escape velocity $\vec{v}_2$, thanks to the exchange of energy and momentum with the more massive PBH of mass $M$.} 
    \label{fig:Slingshot}
\end{figure}

The simplest solution to those problems would be that numerical simulations accounting for baryonic physics explain well the observations~\cite{Sawala:2014xka,2016arXiv160205957W}.  This is however controversial, and possible exotic solutions have been explored, such as dark radiation and interacting dark matter~\cite{Vogelsberger:2015gpr}.   Another solution is that ultra-faint dwarf galaxies exist in large numbers but are now beginning to be found with present sensitivities~\cite{Drlica-Wagner:2015ufc}, thus accounting for the missing satellite population~\cite{Simon:2007dq}.   The scenario of massive PBH-DM could provide an explanation for the existence of those DM-dominated ultra-faint galaxies: dense gas clouds could have ended inside the massive PBH, while stars could have been expelled from the shallow potential wells by fast moving massive PBH through the gravitational slingshot effect. Typical escape velocities of stars in Dwarf Spheroidals (DSph) are tens of km/s. Multiple scatters of stars in massive PBH will increase their speed and eject them from the shallow potential wells. Energy-momentum conservation implies that a star of mass $m$, with incoming velocity $\vec v_1$, that encounters a PBH with mass $M$ and velocity $\vec U$, acquires a velocity $\vec v_2$  through gravitational recoil, see Fig.~\ref{fig:Slingshot}, with magnitude given by
\be\label{recoil}
v_2 = \frac{2U + (1-q)\,v_1}{1+q}\,,
\ee
where $q=m/M$. If the PBH has a large velocity (since it probes the core of the potential well) then it will imprint on the star a velocity twice as large. If the star and the black hole encounter each other at an angle $\theta$, the recoil velocity will be slightly different, 
but marginalizing over angles one gets a mean velocity $\bar v_2 \simeq 2U / (1+q) $ 
which is essentially identical to Eq.~(\ref{recoil}), for $q \ll 1$, and $v_1 \ll U$.

With a few of these BH bypasses (slingshots), the star will acquire a velocity above the escape velocity of the globular cluster or the dwarf spheroidal, and be expelled from the shallow potential well. This may be the reason why small substructures like dwarf galaxy satellites, with masses below $10^6$ to $10^8~M_\odot$, do not shine: they could have lost most of their stars by ejections, and thus present today large mass-to-light ratios, of order 300 to 1000, as recently measured~\cite{Drlica-Wagner:2015ufc}. This may explain both the substructure and the too-big-to-fail problems of standard CDM scenarios.   {In addition, in the case of a relatively broad PBH mass spectrum, a large fraction of the less massive PBH would be ejected from the cluster through the same process, so that those substructures should be today populated only by the most massive ones.   }

%

\textit{Broad PBH mass spectrum:}†   We now focus on the case PBHs follow a broad distribution of masses.  For simplicity we have considered a lognormal distribution of the PBH (local) density, with a central mass $\mu_{\rr{\rm PBH}}$ and a width $\sigma_{\rr{\rm PBH}}$,
\be
\rho(\mPBH)= \frac{\deltaloc \rho_{\rm DM}^0 }{\sqrt{2 \pi \sigma_{\rr{\rm PBH}}^2} } \exp \left[ - \frac{\log^2 (\mPBH / \mu_{\rr{\rm PBH}})}{2 \sigma_{\rr{\rm PBH}}^2} \right]~.
\ee
As before, the local density contrast models how PBHs are clustered, with enhanced local densities.  
Such a distribution is expected e.g. in the scenario of Ref.~\cite{Clesse:2015wea}, in the absence of important merging\footnote{The impact of the merging history on the mass distribution is an important unresolved question that will be considered in a future work.}.  

\begin{figure}
\begin{center}
\includegraphics[scale=1.1]{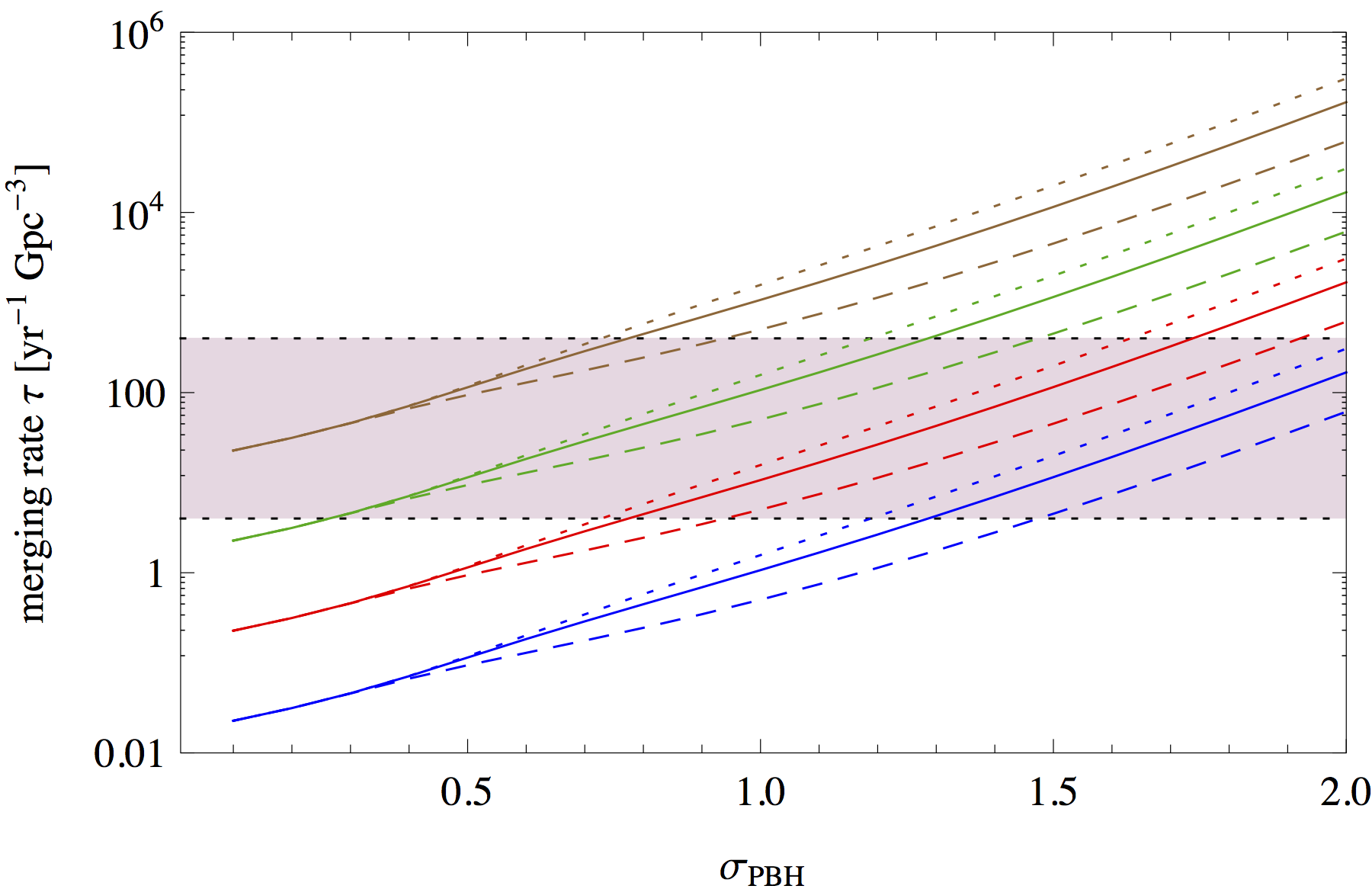}
\caption{Cosmological merging rate as a function of the width $\sigma_{\rr{\rm PBH}} $ of the PBH density spectrum, for different values of the central mass of the distribution $\mu_{\rr{\rm PBH}} = 10 /30 / 60 M_\odot$ (respectively dotted, solid and dashed lines), and of the local density contrast $\deltaloc = 10^7 / 10^8 /10^9 /10^{10}$ (respectively blue, red, green and brown lines).  The colored band corresponds to the bounds inferred by aLIGO.  PBH with broader density spectra or higher masses require less intense clustering but can lead to merging rates within these bounds.  }
\label{fig:broadspectrum}
\end{center}
\end{figure}

The individual merging rate of some PBH of mass $m_{\rr B}$ with any PBH of mass $m_{\rr A}$ in the range $m_{\rm min} < m_{\rr A} < m_{\rr B}$  is given by
\be \label{eq:ind_rate_broad}
\tau(m_{\rr B}) = \int_{m_{\rm min}}^{m_{\rr B}} \tau_{\rr{\rm PBH}}^{\rr{capt}} (m_{\rr B}, m_{\rr A})   \dd(\log m_{\rr A}) 
\ee
where $ \tau_{\rr{\rm PBH}}^{\rr{capt}} (m_{\rr B}, m_{\rr A})  $ is given by Eq.~(\ref{eq:capture_rate}).  The total rate of mergers over some volume ($V=1~\rr{Gpc}^3$) is given by 
\be \label{eq:tot_rate_broad}
\tau_{\rr{\rm PBH}} = \int_{m_{\rm min}}^{m_{\rm max}}  \frac{ \tau(m_{\rr B})  \rho(m_{\rr B}) f_{\rr{DM}} V }{m_{\rr B} \deltaloc } \dd(\log m_{\rr B}).
\ee
 We have calculated and represented on Fig.~\ref{fig:broadspectrum} the merging rate of PBHs with a minimal mass $m_{\rm min}=1 M_\odot$, and a maximal mass $m_{\rm max} = \mu_{\rm PBH} + 10^{3 \sigma_{\rm PBH}}$, as a function of the width $\sigma_{\rr{\rm PBH}} $ of the lognormal PBH density distribution, for several values of the local density contrast and of the central mass $\mu_{\rr{\rm PBH}}$.   Increasing the width of the mass spectrum enhances the merging rate, whereas the local density contrast rescales the merging rate linearly.   In order to accommodate a merging rate ranging from a few to a few hundreds of events per year and per Gpc$^3$, we find that $10^9 \lesssim \deltaloc \lesssim 10^{11}$ if the width of the distribution is negligible, and $10^6 \lesssim \deltaloc \lesssim 10^{8}$ if $\sigma_{\rr{\rm PBH}} \approx 2$.   As expected, in the small width limit the merging rate is independent of $\mu_{\rr{\rm PBH}}$.   But we find that for $\sigma_{\rr{\rm PBH}} \gtrsim 0.5$, the merging rate becomes mass-dependent and is more important for large values of $\mu_{\rr{\rm PBH}}$.    These results assume a constant density inside PBH clusters, but a refined treatment should include particular profiles for the PBH distribution inside sub-halos, which would be obtained by convolving Eq.~(\ref{eq:tot_rate_broad})  with some motivated profile for DM sub-halos~\cite{1989A&A...223...89E}, like Einasto's one~\cite{2013MNRAS.431.1220D}.  
 
\begin{figure*}
\begin{center}
\includegraphics[scale=0.85]{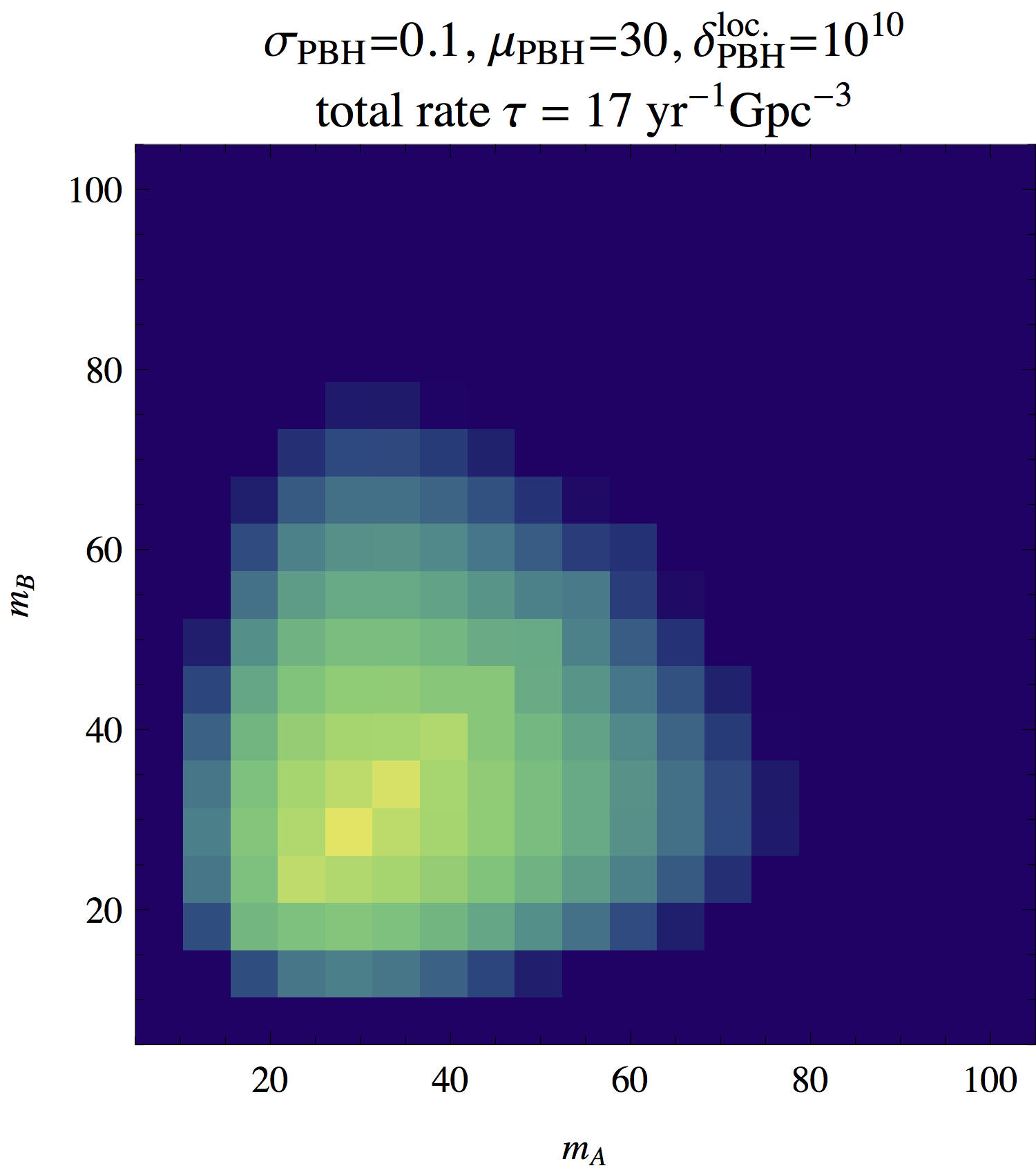}  \includegraphics[scale=0.75]{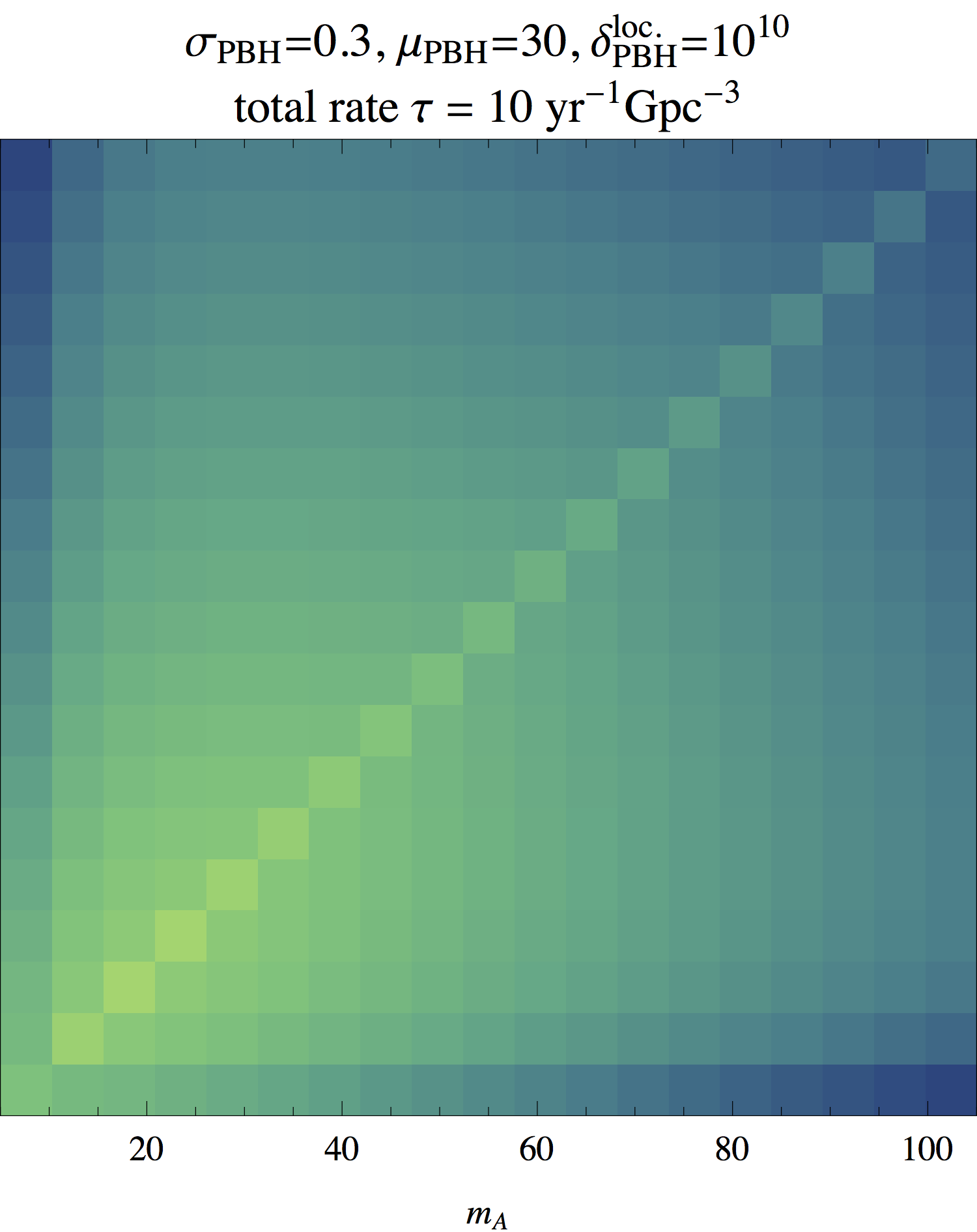}
\includegraphics[scale=0.75]{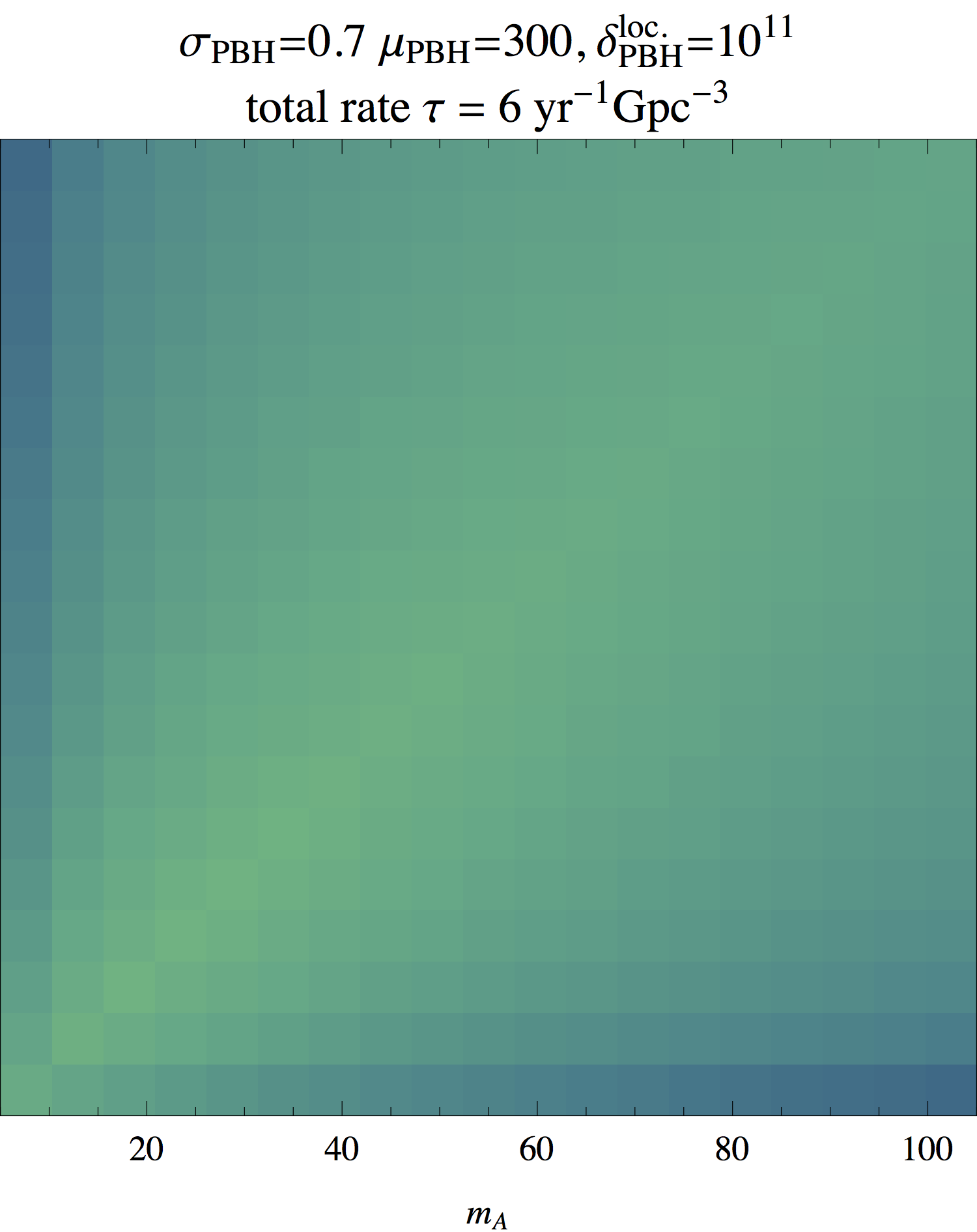} \includegraphics[scale=0.7, trim=0 -1.2cm 0 0]{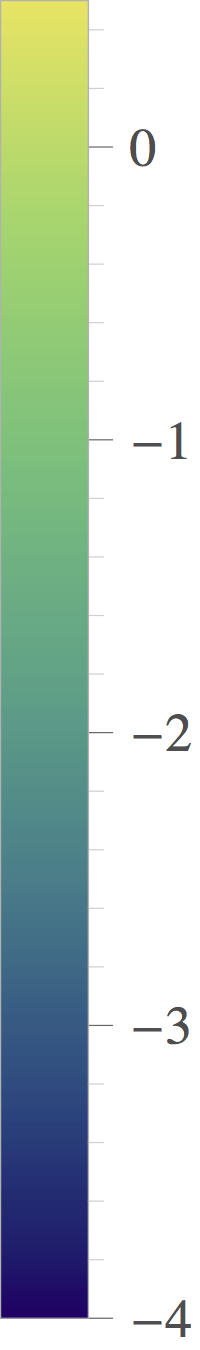}
\caption{Cosmological merging rates of BHs with masses $m_{\rr A}$ and $m_{\rr B}$, the color scale representing $\log (\tau~\rr{yr}~\rr{Gpc}^3 )$.   The PBH density follows a lognormal distribution 
in logarithmic mass scale, of central value $\mu_{\rr{\rm PBH}}$ and width $\sigma_{\rr{\rm PBH}}$.  The panels are for $\mu_{\rr{\rm PBH}} = 30 M_\odot$ and $\sigma_{\rr{\rm PBH}} = 0.1$ ({1st panel}), $\sigma_{\rr{\rm PBH}} = 0.3$ ({2nd}),  
and $\mu_{\rr{\rm PBH}} = 300 M_\odot$ {with  $\sigma_{\rr{\rm PBH}} = 0.7$}  {(3rd)}
The corresponding local density contrast and total merging rate of BHs with masses $\mPBH \gtrsim 5 M_\odot$ is indicated above each panel. Those total rates lie all within the bounds inferred by aLIGO, but with different distributions in the $(m_A, m_B)$ plane, which is potentially detectable and makes possible to reconstruct the PBH density spectrum.  }
\label{fig:PBHclusters}
\end{center}
\end{figure*}

One can also study how the merging events are distributed as a function of the two progenitor masses $m_A$ and $m_B$.  Fig.~\ref{fig:PBHclusters} displays the merging rate for $m_A$ and $m_B$ ranging from $5 M_\odot$ to $100 M_\odot$, with mass bins of width $5 M_\sun$.  With a nominal PBH mass $\mu_{\rm PBH} = 30 M_\odot$ and a small width $\sigma_{\rm PBH}  \simeq 0.1$ events involving one PBH with a mass difference of $\pm 10 M_\odot$ are very unlikely.   On the other hand, with a width $\sigma_{\rm PBH}  \simeq 0.3$ it would be possible to observe events involving two PBHs with a big mass difference.   { Increasing further the width to $\sigma_{\rm PBH} \gtrsim 0.5$ and the most likely events then come from the less massive PBHs (of a few solar masses), because the number density of PBHs $n_{\rm PBH} \equiv \rho_{\rm PBH} / \mPBH$ peaks on those scales.   One can nevertheless produce comparable merging rates to aLIGO by considering a broad PBH distribution centered on higher masses together with a higher density contrast.   But such a scenario is limited because of the current constraints from the disruption of wide binaries that do not allow a large fraction of the DM to be made of very massive (thousands of solar masses) PBHs.  }


%


These considerations lead us to the conclusion that if Advanced LIGO or other gravitational wave experiments detect, within the next few years, a large number  ($\gtrsim 1000$) of merging events involving massive BHs, we should be able to reconstruct the PBH mass spectrum (with $\sim 10$\% accuracy), as well as their possible local density environments and the fraction of dark matter they account for.  

\textit{Summary and discussion:}   The detection by aLIGO of gravitational waves emitted by the merging of two massive BHs opens a new way to probe the abundance, the clustering and the mass distribution of PBHs.   The merging rates expected for various local densities and mass distributions have been calculated and compared to the bounds $2 - 400$ Gpc$^{-3}$ yr$^{-1}$  inferred by aLIGO, in the case PBHs have the right abundance for being the dark matter.   A uniform distribution of PBHs inside galactic halos cannot reproduce such high rates.  But we find that if PBHs are clustered in sub-halos with densities comparable to the one of DM-dominated ultra-faint dwarf galaxies, the merging rate lies precisely within that range.  We suggest a model where PBHs are massive -- a few tens of solar masses -- and have a broad mass spectrum, like the one generated by hybrid inflation with a mild waterfall phase~\cite{Clesse:2015wea},  such that a subdominant number of very massive PBHs can be the seeds of the SMBH at the center of galaxies, as well as of the IMBHs expected to be at the origin of ultra-luminous X-ray sources.  Such a PBH-DM model would have interesting observational consequences and could solve the long-standing \textit{missing satellite} and \textit{too-big-to-fail} problems of $\Lambda$CDM cosmology. Finally, by studying the merging rates with different progenitor masses in the range $5 - 100 M_\odot$, we find that the detection of thousands of merging events by LIGO, VIRGO and future GW detectors like KAGRA, would allow the reconstruction the PBH mass spectrum with relatively good accuracy.   

The detailed study of the merging history and its impact on the PBH mass spectrum is left for future work, as well as the calculation of the associated stochastic background of gravitational waves, which could reach a non-negligible fraction of the critical density and lead to observable signatures in the CMB and LSS.   Moreover, constraints from the absence of microlensing of stars in the Magellanic clouds are evaded due to the large mass of PBHs.  Furthermore, if PBHs are clustered in dwarf satellite galaxies, then the probability of finding one of them in the line of sight of the Large Magellanic Cloud is less than a part in a thousand, and thus we argue that a model where an important fraction of PBHs have a sub-stellar mass is still allowed.  Finally, the GAIA experiment should set new bounds on PBH abundances from anomalous motions of stars, and should be able to distinguish between the different cases: clustered/unclustered, sharp/broad mass spectrum.  This line of research will be pursued further in future work.      

\textit{Acknowledgments:}  
This work is supported by the Research Project of the Spanish MINECO, FPA2013-47986-03-3P, and the Centro de Excelencia Severo Ochoa Program SEV-2012-0249. JGB and SC thank the CERN Theory Division for its kind hospitality during the summer of 2015, when part of this work was initiated.

\bibliography{biblio}

\end{document}